\documentstyle[eqsecnum,aps,epsf]{revtex}

\begin{document}
\draft
\title{Ground state ferromagnetism in a doubly orbitally degenerate model}
\author{L.~Didukh\cite{e-mail}, Yu.~Skorenkyy,V.~Hankevych, and O.~Kramar}
\address{Ternopil State Technical University, Department of Physics,\\
56 Rus'ka Str., Ternopil UA-46001, Ukraine }
\date{\today}

\maketitle

\begin{abstract}
In the present paper the ground state of a double orbitally degenerate
model at weak intra-atomic interaction is studied using the Green
functions method. Beside the diagonal matrix elements of electron-electron
interactions the model includes correlated hopping integrals and 
inter-atomic exchange interaction. The influence of orbital degeneracy
with Hund's rule coupling,
correlated hopping and inter-atomic direct exchange on the ferromagnetic 
ordering is investigated. The expressions for ground state energy and 
magnetization, the criterion of transition from paramagnetic to ferromagnetic 
ground state as functions of the model parameters are obtained.
The obtained results are compared with some experimental data for
magnetic materials.
\end{abstract}

\section{Introduction}

The problem of an origin of the metallic
ferromagnetism, in spite of the variety of theoretical attempts to solve it,
still remains open. Nowadays we can distinguish a few ways to obtain the 
ferromagnetic solution. First, one 
can consider the Hubbard model~\cite{hub}-\cite{gutz} which describes 
itinerant 
electrons in a single non-degenerate band interacting via on-site Coulomb 
repulsion $U$. Within this model the ferromagnetic solution has been obtained
by means of some approximations~\cite{dieter,nolt}, also the exact result of
Nagaoka~\cite{nag} shows that the 
ground state of a band with exactly one electron above or below 
half-filling is 
ferromagnetic at $U=\infty$. However, in spite of the large number of 
papers (for recent reviews see Refs.~\cite{dieter,tas})
the question of an existence of ferromagnetic ordering in the Hubbard model
is still under discussion.
Second, one can include in a model Hamiltonian, in addition to the
intra-atomic Coulomb repulsion, also other matrix elements of 
electron-electron 
interaction, which can provide new mechanisms of ferromagnetism 
stabilization. Third, one can take into consideration the orbital 
degeneracy with the intra-atomic Hund's rule exchange interaction which 
forms atomic magnetic moments. Taking into consideration the orbital 
degeneracy is essentially important: in the case of strong intra-atomic 
iteractions the kinetic superexchange has a ferromagnetic nature, in
contrast with the single-band Hubbard model where it has an antiferromagnetic
one. We believe that at least for understanding of the 
ferromagnetism origin in crystals with narrow energy bands it is enough to
consider other matrix elements of electron correlations (in addition to the 
intra-atomic Coulomb
repulsion) and orbital degeneracy with intra-atomic Hund's rule exchange.

The importance of orbital degeneracy and Hund's rule exchange interaction
for ferromagnetism  was first suggested by Slater~\cite{sl} and van 
Vleck~\cite{vl}: the presence of orbital degeneracy and Hund's rule exchange
interaction leads to so-called ``atomic ferromagnetism''; in such a situation
translational motion of electrons forces the spins of electrons on 
nearest-neighbour atoms to align in parallel. This mechanism of 
ferromagnetism based on a microscopic model was studied by Roth~\cite{roth}
considering the Hubbard model with double orbital degeneration. In 
Refs.~\cite{kk,cl,gs,ks} it was found that ferromagnetism in such two-band
Hubbard model coexists with orbital ordering for the case of a 
quarter-filled band and strong intra-atomic interactions. Moreover, recently
Spa\l ek with co-workers~\cite{klej} have proposed the mechanism of a coexisting 
ferromagnetism and spin-triplet paired state in a doubly orbitally degenerate
band due to the intra-atomic Hund's rule coupling. The problem of metallic 
ferromagnetism in the two-band Hubbard model has attracted much attention
of researchers 
in a series of papers by means of the dynamical mean-field 
theory~\cite{voll,vollh,mk}, the slave boson method~\cite{klej,fk} and
the Gutzwiller variational wave function approximation~\cite{chao,bw,bw1,oka}.
They all find ferromagnetism to be stabilized by intra-atomic Hund's rule
coupling at intermediate and strong intra-atomic Coulomb interactions. On
the other hand, the authors~\cite{hir97,mn} have obtained that intra-atomic
Hund's rule exchange does not play the important role in ferromagnetic
ordering of crystals with narrow energy bands, and even may destabilize 
ferromagnetism. In particular, Hirsch~\cite{hir97} suggests that orbital
degeneracy and intra-atomic exchange interaction are not likely to play
a significant role in the ferromagnetic ordering of Ni or Ni-Cu and
Ni-Zn alloys, and a single-band model with inter-atomic direct exchange 
contains the essential physics of metallic ferromagnetism of transition 
metal compounds. Nolting and co-workers~\cite{mn} have obtained that
magnetization of the doubly orbitally degenerate Hubbard model strongly
decreases with increasing intra-atomic Hund's rule exchange for a wide 
range of model parameters, namely the intra-atomic exchange coupling
substantially suppresses ferromagnetic order.

In the cited above papers the authors have not took into consideration
so-called ``off-diagonal'' matrix elements of electron-electron interaction 
which are one of the possible mechanisms of ferromagnetism in
narrow energy bands as mentioned above. The importance of these matrix 
elements in a 
single-band case was pointed out in many 
works~\cite{dieter,voll,did}-\cite{koll}. Here
we note the special role of direct exchange interaction 
and correlated hopping (taking into account of the inter-atomic
density-density Coulomb interaction which plays the essential role in charge
ordering goes beyond the goal of this article).
Last ten years the problem of importance of inter-atomic exchange interaction
for the metallic ferromagnetism again is under discussion in number of 
works~\cite{dieter,voll,did}-\cite{wahl}, 
where it was concluded that the inter-atomic exchange 
interaction $J$ plays a fundamental role for the stabilization of 
ferromagnetic ordering in a single-band model. The authors found that
the inter-atomic direct exchange $J>0$ stabilizes ferromagnetic ordering
in the single-band Hubbard model at intermediate to strong intra-atomic
Coulomb interactions~\cite{dieter,voll,amad,wahl}. In the case of weak 
interactions the inter-atomic exchange is important for the stabilization
of incomplete ferromagnetism~\cite{cm00,hir_t0}. Hirsch~\cite{hir_t0}
argued that the inter-atomic exchange interaction is the main driving
force for metallic ferromagnetism in systems like iron, cobalt and nickel.
Also for the special cases of generalized Hubbard models it has been 
found~\cite{Str&Voll,ksv} by means of exact techniques that inter-atomic
exchange plays a dominant role in the occurrence of ferromagnetism.

The importance of correlated hopping for understanding of the metallic
ferromagnetism in narrow energy band was discussed in 
Refs.~\cite{voll,did,cm00,kivel,camp,amad,Str&Voll,ksv,koll}. In particular,
a generalization of Nagaoka's theorem has been proved~\cite{ksv}, 
and it has been shown~\cite{dps,amad} that in strong coupling regime
close to half-filling correlated hopping favours ferromagnetism stronger 
for electron-like carriers than for hole-like carriers
(the reverse situation occurs at weak interactions~\cite{amad}).
Note, that due to an additional mechanism of correlated 
hopping~\cite{cm00} at weak intra-atomic interactions the situation, which
is analogous to that of strong interactions, can be realized: correlated 
hopping favours ferromagnetism stronger for electron-like carriers
versus hole-like carriers (see also Section~3).

In this connection, the necessity of a further study of the metallic
ferromagnetism problem in narrow energy bands is obvious. Firstly,
it is interesting and important to find how 
the orbital degeneracy with intra-atomic Hund's rule coupling and 
``off-diagonal'' matrix elements of electron-electron interaction 
(correlated hopping and inter-atomic direct exchange interaction) 
in the aggregate show itself. Note
that these studies were performed partially in Ref.~\cite{hir97} by means of
exact diagonalization, in particular, for the case of small one-dimensional 
chains and strong intra-atomic Coulomb repulsion the role of orbital 
degeneracy with intra-atomic Hund's rule coupling and inter-atomic
exchange interaction for the ferromagnetic ordering was studied. However,
the results depend sensitively on the number of lattice sites and the
boundary conditions, on the one hand, and the study is restricted to the
one-dimensional case, on the other hand. Secondly, there is a contradiction
about the role of intra-atomic Hund's rule interaction for the stabilization
of ferromagnetic ordering, as mentioned above. Therefore, the present paper
is devoted to the study of the metallic ferromagnetism problem.

The structure of the paper is the following. In Section~2 we formulate the
Hamiltonian of the doubly orbitally degenerate Hubbard model which is 
generalized by taking into account correlated hopping and inter-atomic
exchange interaction. For the case of weak intra-atomic interactions the
single-particle Green function and energy spectrum 
at arbitrary values of electron concentration are derived by means of the 
mean-field approximation. In Section~3 the ferromagnetism in ground state of
the model is investigated. The role of orbital degeneracy with intra-atomic
Hund's rule coupling, of correlated hopping, and of inter-atomic direct
exchange interaction for the stability of ferromagnetic ordering is studied.
The expressions for ground state energy and magnetization as functions of  
the model parameters, the criterion of transition from paramagnetic
to feromagnetic ground state are found. Finally, Section~4 is devoted to
the conclusions from the obtained results.

\section{Green function and energy spectrum of the model in the case of 
weak interaction}

Let us generalize the Hamiltonian proposed in work~\cite{prb}
by taking into account the inter-atomic exchange interaction:

\begin{eqnarray} \label{H1}
H=&&-\mu \sum_{i\gamma\sigma}a_{i\gamma\sigma}^{+}a_{i\gamma\sigma}+
{\sum\limits_{ij\gamma\sigma}}'t_{ij}(n) a_{i\gamma\sigma}^{+}
a_{j\gamma\sigma}+{\sum\limits_{ij\gamma\sigma}}'(t'_{ij}
a_{i\gamma\sigma}^{+}a_{j\gamma\sigma}n_{i\bar{\gamma}} + h.c.)
\nonumber\\
&&+{\sum\limits_{ij\gamma\sigma}}'(t''_{ij}
a_{i\gamma\sigma}^{+}a_{j\gamma\sigma}n_{i\gamma\bar{\sigma}} + h.c.)
+U \sum_{i\gamma} n_{i\gamma\uparrow}n_{i\gamma\downarrow}
+U' \sum_{i\sigma}n_{i\alpha\sigma}n_{i\beta\bar{\sigma}}
\\
&&+(U'-J_0)\sum_{i\sigma}n_{i\alpha\sigma}n_{i\beta\sigma}
\nonumber
+J_0\sum_{i\sigma} a_{i\alpha\sigma}^{+}a_{i\beta\bar{\sigma}}^{+}
a_{i\alpha\bar{\sigma}}a_{i\beta\sigma}
\\
&&+{J\over 2}{\sum_{ij \gamma \gamma ' \sigma \sigma^{'}}}' 
a_{i\gamma\sigma}^{+}a_{j\gamma{'}\sigma{'}}^{+}
a_{i\gamma\sigma{'}}a_{j\gamma{'}\sigma},
\nonumber
\end{eqnarray}
where 
$\mu$ is the chemical potential, $a_{i\gamma\sigma}^{+}, a_{i\gamma\sigma}$ 
are the creation and destruction 
operators of an electron of spin $\sigma$ ($\sigma =\uparrow, \downarrow$;
${\bar \sigma}$ denotes spin projection which is opposite to $\sigma$) 
on $i$-site and in orbital $\gamma$ ($\gamma=\alpha ,\beta$ denotes two 
possible orbital states),  
$n_{i\gamma\sigma}=a_{i\gamma\sigma}^{+}a_{i\gamma\sigma}$
is the number operator of electrons of spin $\sigma$ and in orbital $\gamma$ 
on $i$-site, $n_{i\gamma}=n_{i\gamma\uparrow}+n_{i\gamma\downarrow}$;
$t_{ij}(n)$ is the effective concentration-dependent hopping integral of 
an electron
from $\gamma$-orbital of $j$-site to $\gamma$-orbital of $i$-site  
(we neglect the electron hoppings between $\alpha$- and $\beta$-orbitals), 
$t'_{ij}$ ($t''_{ij}$) includes influence of an electron
on $\bar{\gamma}$ ($\gamma$)-orbital of $i$- or $j$-site 
on hopping process 
($\bar{\gamma}=\beta$ if $\gamma=\alpha$, and $\bar{\gamma}=\alpha$ when
$\gamma=\beta$),
the primes at sums in Eq.~(\ref{H1}) signify that $i\not=j$,
$U$
is the intra-atomic Coulomb repulsion of two electrons of the opposite spins
at the same orbital (we assume that it has the same value at $\alpha$- and 
$\beta$-orbitals),
$U'$
is the intra-atomic Coulomb repulsion of two electrons of the opposite spins
at the different orbitals,
$J_0$
is the intra-atomic exchange interaction energy which stabilizes the Hund's
states forming the atomic magnetic moments,
and $J$
is the inter-atomic exchange interaction.
The effective hopping integral $t_{ij}(n)$ is  
concentration-dependent in consequence of taking into account the correlated
hopping~\cite{cmp} of electron.

The peculiarities of the model described by the Hamiltonian~(\ref{H1}) 
are taking into consideration the influence of the site occupation on the 
electron hoppings (correlated hopping), and the direct exchange 
between the neighbouring sites.
In this model an electron hopping from one site to another is correlated 
both by the occupation
of the sites involved in the hopping process and the 
occupation of the nearest-neighbour sites.
The correlated hopping, firstly, renormalizes the initial hopping 
integral (it becomes concentration- and spin-dependent) and, secondly,
leads to an independent on quasiimpulse shift of the subband center, dependent 
on magnetic and orbital orderings. 
To characterize the value of correlated hopping we introduce dimensionless
parameters  $\tau={t_{ij}(n) \over |t_{ij}|}-1$, 
$\tau'={{t'}_{ij} \over |t_{ij}|}$, and $\tau_2={{t''}_{ij}\over |t_{ij}|}$
where $t_{ij}$ is the band hopping integral.

The single-particle Green function satisfies the equation
\begin{eqnarray}
&&(E+\mu)\langle\langle a_{p\gamma\sigma}
|a^{+}_{p'\gamma\sigma}\rangle\rangle_{E}
=\frac{\delta_{pp'}}{2\pi}
+\sum_{i}t_{ip}(n)\langle\langle a_{i\gamma\sigma}
|a^{+}_{p'\gamma\sigma}\rangle\rangle_{E}
\\ 
\nonumber
&&+\sum_{i\sigma'}t'(ip)\Bigl[\langle\langle a^{+}_{p\bar{\gamma}\sigma'}
a_{p\bar{\gamma}\sigma'}a_{i\gamma\sigma}
|a^{+}_{p'\gamma\sigma}\rangle\rangle_{E}
+\langle\langle a^{+}_{p\bar{\gamma}\sigma'}
a_{i\bar{\gamma}\sigma'}a_{p\gamma\sigma}
|a^{+}_{p'\gamma\sigma}\rangle\rangle_{E}
\\
\nonumber
&&+\langle\langle a^{+}_{i\bar{\gamma}\sigma'}
a_{i\bar{\gamma}\sigma'}a_{i\gamma\sigma}
|a^{+}_{p'\gamma\sigma}\rangle\rangle_{E}
+\langle\langle a^{+}_{i\bar{\gamma}\sigma'}
a_{p\bar{\gamma}\sigma'}a_{p\gamma\sigma}
|a^{+}_{p'\gamma\sigma}\rangle\rangle_{E}\Bigr]
\\ 
\nonumber
&&+\sum_{i}t''(ip)\Bigl[\langle\langle a^{+}_{p\gamma\bar{\sigma}}
a_{p\gamma\bar{\sigma}}a_{i\gamma\sigma}
|a^{+}_{p'\gamma\sigma}\rangle\rangle_{E}
+\langle\langle a^{+}_{p\gamma\bar{\sigma}}
a_{i\gamma\bar{\sigma}}a_{p\gamma\sigma}
|a^{+}_{p'\gamma\sigma}\rangle\rangle_{E}
\\
\nonumber
&&+\langle\langle a^{+}_{i\gamma\bar{\sigma}}
a_{i\gamma\bar{\sigma}}a_{i\gamma\sigma}
|a^{+}_{p'\gamma\sigma}\rangle\rangle_{E}
+\langle\langle a^{+}_{i\gamma\bar{\sigma}}
a_{p\gamma\bar{\sigma}}a_{p\gamma\sigma}
|a^{+}_{p'\gamma\sigma}\rangle\rangle_{E}\Bigr]
\\ 
\nonumber
&&
+U\langle\langle n_{p\gamma\bar{\sigma}}a_{p\gamma\sigma}
|a^{+}_{p'\gamma\sigma}\rangle\rangle_{E}
+U'\langle\langle n_{p\bar{\gamma}\bar{\sigma}}a_{p\gamma\sigma}
|a^{+}_{p'\gamma\sigma}\rangle\rangle_{E}
+(U'-J_0)\langle\langle n_{p\bar{\gamma}\sigma}a_{p\gamma\sigma}
|a^{+}_{p'\gamma\sigma}\rangle\rangle_{E}
\\ 
\nonumber
&&
+J_0\langle\langle a^{+}_{p\bar{\gamma}\bar{\sigma}}
a_{p\gamma\bar{\sigma}}a_{p\bar{\gamma}\sigma}
|a^{+}_{p'\gamma\sigma}\rangle\rangle_{E}
+\sum_{i\gamma'\sigma'} J
\langle\langle a^{+}_{i\gamma'\sigma'}a_{p\gamma\sigma'}
a_{i\gamma'\sigma}
|a^{+}_{p'\gamma\sigma}\rangle\rangle_{E}.
\nonumber
\end{eqnarray}

Let us consider the system at weak intra-atomic Coulomb interaction 
($U$, $U'$ are smaller than the bandwidth $w=2z|t_{ij}|$ where
$z$ is the number of nearest neighbours to a site). 
In this case we can take into account electron-electron 
interactions in the Hartree-Fock approximation:
\begin{eqnarray}\label{appr}
\langle\langle a^{+}_{i\gamma\bar\sigma}a_{i\gamma\bar\sigma}
a_{j\gamma\sigma}|a^{+}_{p'\gamma\sigma}\rangle\rangle_{E}
&\simeq&\langle a^{+}_{i\gamma\bar\sigma}a_{i\gamma\bar\sigma}\rangle
\langle\langle a_{j\gamma\sigma}|a^{+}_{p'\gamma\sigma}\rangle\rangle_{E};\\
\nonumber
\langle\langle a^{+}_{i\gamma\bar\sigma}a_{j\gamma\bar\sigma}
a_{i\gamma\sigma}|a^{+}_{p'\gamma\sigma}\rangle\rangle_{E}
&\simeq&\langle a^{+}_{i\gamma\bar\sigma}a_{j\gamma\bar\sigma}\rangle
\langle\langle a_{i\gamma\sigma}|a^{+}_{p'\gamma\sigma}\rangle\rangle_{E}.
\end{eqnarray}
We assume that averages 
$\langle a^{+}_{i\gamma\sigma}a_{i\gamma\sigma}\rangle=n_{\gamma\sigma}$
are independent of the number of a site, i.e. we limit ourselves to a 
uniform charge and electronic magnetic moment distribution (the problems of
antiferromagnetic, orbital and charge orderings will be studied elsewhere).

After the transition to Fourier representation we obtain for the Green 
function 
\begin{eqnarray}\label{grf}
&& \langle\langle a_{p\gamma\sigma}|a^{+}_{p'\gamma\sigma}\rangle\rangle_{\bf k}
=\frac{1}{2\pi}\frac{1}{E-E_{\gamma\sigma}(\bf k)},
\end{eqnarray}
where the single particle energy spectrum is
\begin{eqnarray}\label{spectr}
E_{\gamma\sigma}({\bf k})=-\mu_{\gamma\sigma}+t_{\bf k}(n\gamma\sigma),
\end{eqnarray}
with the shifted chemical potential 
\begin{eqnarray}
\mu_{\gamma\sigma}=\mu-{\beta'}_{\gamma}-{\beta''}_{\gamma\sigma}
-n_{\gamma\bar{\sigma}}U
-n_{\bar{\gamma}\bar{\sigma}}U'-n_{\bar{\gamma}\sigma}(U'-J_0)
+zJ\sum_{\sigma'}n_{\gamma\sigma'},
\end{eqnarray}
here the shifts of the subband centers are
\begin{eqnarray}
&& {\beta'}_{\gamma}={2 \over N}\sum_{ij\sigma} t'(ij) \langle 
a_{i\bar{\gamma}\sigma}^{+}a_{j\bar{\gamma}\sigma}\rangle,
\\
&& {\beta''}_{\gamma\sigma}={2 \over N}\sum_{ij} t''(ij) \langle 
a_{i\gamma\bar{\sigma}}^{+}a_{j\gamma\bar{\sigma}}\rangle;
\end{eqnarray}
and the spin- and concentration-dependent hopping integral is
\begin{eqnarray}
 t_{\bf k}(n\gamma\sigma)=
t_{\bf k}(1-\tau n-2\tau'n_{\bar{\gamma}}
-2\tau_2 n_{\gamma\bar{\sigma}}-{zJ\over w} \sum_{\sigma'}
\langle a_{i\gamma\sigma'}^{+}a_{j\gamma\sigma'}\rangle),
\end{eqnarray}
$t_{\bf k}$ is the Fourier transformant of the hopping integral $t_{ij}$. 

 The dependence of effective hopping integral on electron concentration 
and magnetization, a presence of the spin-dependent shift of subband
center are the essential distinctions of single-particle energy spectrum
of the model described by Hamiltonian~(\ref{H1}) from the spectrum of the 
Hubbard model in the case of weak interaction.

\section{Ferromagnetism in the ground state of the model}

The concentration of electrons with spin $\sigma$ on $\gamma$ orbital is 
\begin{eqnarray}
n_{\gamma\sigma}=
\int\limits_{-\infty}^{+\infty}\rho(\epsilon)
f(E_{\gamma\sigma}(\epsilon))d\epsilon.
\end{eqnarray}
Here $\rho(\epsilon)$ is the density of states, $f(\epsilon)$ is the Fermi 
distribution function, $E_{\gamma\sigma}(\epsilon)$ is obtained from
respective formula~(\ref{spectr}) substituting 
$t_{\bf k}\!\rightarrow\! \epsilon$.
Let us assume the rectangular density of states:
\begin{eqnarray}
\rho(\epsilon)=\frac{1}{N}\sum_{\bf k}\delta(\epsilon-\epsilon({\bf k}))=
\frac{1}{2w}\theta(\epsilon^{2}-w^{2}).
\end{eqnarray}

In the case of zero temperature we obtain:
\begin{eqnarray}\label{nsigma}
n_{\gamma\sigma}=\frac{\varepsilon_{\gamma\sigma}+w}{2w},
\end{eqnarray}
where the value
$\varepsilon_{\gamma\sigma}$ is the solution of the equation 
$E_{\gamma\sigma}(\varepsilon)=0$,
from which we obtain
$\varepsilon_{\gamma\sigma}=\frac{\mu_{\gamma\sigma}}{\alpha_{\gamma\sigma}}$,
where
$\alpha_{\gamma\sigma}=1-\tau n-2\tau'n_{\gamma}
-2\tau_2 n_{\gamma\bar{\sigma}}
-{zJ\over w}\sum\limits_{\sigma'}n_{\gamma\sigma'}(1-n_{\gamma\sigma'})$.

The shifts of subband centers are
\begin{eqnarray}
&&{\beta'}_{\gamma}={2 \over N}\sum_{ij\sigma} t'(ij) \langle 
a_{i\bar{\gamma}\sigma}^{+}a_{j\bar{\gamma}\sigma}\rangle=
-2\tau' w\sum_{\sigma}n_{\bar{\gamma}\sigma}(n_{\bar{\gamma}\sigma}-1),
\\
&& {\beta''}_{\gamma\sigma}={2 \over N}\sum_{ij} t''(ij) \langle 
a_{i\gamma\bar{\sigma}}^{+}a_{j\gamma\bar{\sigma}}\rangle=
-2\tau_2 w n_{\gamma\bar{\sigma}}(n_{\gamma\bar{\sigma}}-1).
\end{eqnarray}
From equation~(\ref{nsigma}) we obtain for the magnetization ($m<n$):
\begin{eqnarray}
\label{m0} 
m&=&\sum_{\gamma}(n_{\gamma\uparrow}-n_{\gamma\downarrow})
\nonumber
\\
&=&\pm 2{\left( {zJ \over 2w}\right)}^{- {1 \over 2}}
\left({zJ\over 8w}(8+n(4-n))+\frac{U+J_0}{2w}+\tau_1 n
+{\tau_2\over 2} (4-n)-1\right)^{ 1 \over 2},
\end{eqnarray}
where $\tau_1=\tau +\tau'$, here we have assumed that the orbital distribution of 
electrons is uniform. Note that magnetization~(\ref{m0}) does not 
depend on the parameter of intra-atomic Coulomb interaction $U'$, which
leads to the independent on magnetic moment renormalization of the
chemical potential (it can be seen also from the expression for ground state
energy). 
Similarly, in the work~\cite{mn} is has been argued that this parameter 
does not play the decisive role in metallic ferromagnetism of the 
transition metal compounds.

The magnetization defined by Eq.~(\ref{m0}) is plotted in Fig.~1 as a 
function of 
electron concentration $n$ at different values ${J_0/w}$. These dependencies
qualitatively agree with results of work~\cite{oka} obtained by use of
the Gutzwiller variational functions method.
From Fig.~1 one can see that nature of the ground state of the system 
strongly depends on the values of system parameters; the small changes 
of $J_0$ can lead to the transition from a paramagnetic state
to a ferromagnetic one
at some values of electron concentration and energy parameters
(this result agrees with the results of works~\cite{hir97,vollh}); 
note that at some values of parameters the system can be fully
polarized. The transition to ferromagnetic state is also possible with the 
increase of $n$.
Similar transition with the increase of electron concentration 
has been found by the authors of work~\cite{acq}.

Taking into account correlated hopping leads to the appearance
of a peculiar kinetic mechanism of ferromagnetic ordering stabilization. 
This mechanism is caused by the presence of the spin-dependent shift of
the subband centers being the consequence of correlated hopping
(which are similar to the shift of subband centers in consequence of 
inter-atomic direct exchange interaction).

The influence of correlated hopping on behaviour of the system is 
illustrated on Fig.~2. In distinction from the two-band  Hubbard model
there is an asymmetry of the cases $n<2$ and $n>2$.
With the increase of parameter $\tau_1$ the region of ferromagnetic ordering 
moves towards larger values of electron concentration $n$, and with increasing
$\tau_1$ -- to smaller values of $n$. Let us also note that taking into 
account the correlated hopping significantly enriches the set of curves
(illustrating the $m(n)$ dependencies), which qualitatively describe 
the experimental Slater-Pauling curves~\cite{gau} for ferromagnetic alloys.

The peculiarity of degenerate band models is taking into account 
Hund's exchange interaction $J_0$. The importance of $J_0$ is shown in
Fig.~3. One can see that intra-atomic exchange stabilizes ferromagnetism
in orbitally degenerate band (the behaviour of $m$ with the increase of $J_0$ 
qualitatively agrees with the dependence of magnetic moment on the
intra-atomic correlation strength obtained in work~\cite{acq}). 
To describe the real narrow-band materials we have to take into account 
the correlated hopping which allows to obtain the transition from 
paramagnetic to ferromagnetic phase at realistic values of $J_0$.
Fig.~4, which is plotted with use of Eq.~(\ref{m0}) at ${U \over w}=1.2$, 
${zJ \over w}=0.06$ and ${J_0 \over w}=0.2$, $\tau_1=0$, $\tau_2=0.15$, 
reproduces the behaviour of the magnetization observed in the systems
Fe$_{1-x}$Co$_x$S$_2$ and Co$_{1-x}$Ni$_x$S$_2$ with the change of 
electron concentration in $3d$-band~\cite{jarr}. In these crystals
the same subsystem of electrons is responsible both for conductivity
and for the localized magnetic moment formation.
The noted compounds have the cubic pyrite structure, 
then $3d$-band
is split into two subbands: a doubly degenerate $e_g$ band and
a triply degenerate $t_{2g}$ band; $t_{2g}$ band is completely filled 
and $e_g$ band is partially filled (the $e_g$ band filling changes
from 0 to 1 in the compound Fe$_{1-x}$Co$_x$S$_2$ and from 1 to 2 in 
the compound
Co$_{1-x}$Ni$_x$S$_2$). One should describe $e_g$ band of these 
compounds by a doubly orbitally degenerate model.

The unusual peculiarity of the system Fe$_{1-x}$Co$_x$S$_2$ is the 
presense of ferromagnetic ordering at very small values of electron 
concentration $n=x\simeq 0.05$~\cite{jarr}. Ferromagnetism in this compound
has been studied within a single-band model in Refs.~\cite{aik,did1}.
The authors of work~\cite{aik} have proposed an approximation for the 
description of Fe$_{1-x}$Co$_x$S$_2$ in the non-degenerate Hubbard model
with $U=\infty$ which provides the ferromagnetic solution even at very
small electron concentration (in this connection see also Ref.~\cite{did1}).
However, in accordance with the
Kanamori theory~\cite{kan} at very small $n$ we should obtain the gas limit 
where ferromagnetism does not occur. We also believe that
the degeneracy of $e_g$ band is essential for the description of 
ferromagnetic ordering in this system. 
Our results allow  to obtain the ferromagnetism for small 
values of electron concentration induced by correlated hopping $\tau_2$ 
in a presence of the inter-atomic exchange interaction (see Fig.~4). 
Thus, we believe that the correlated hopping mechanism
in a presence of the inter-atomic exchange interaction allows 
the more natural explanation of the origin of ferromagnetism in the system 
Fe$_{1-x}$Co$_x$S$_2$ at very small $x$.

To calculate the ground state energy of the model per site we use the formula
\begin{eqnarray}\label{E0} 
{E_0}={1\over 2N} \sum_{{\bf k}\gamma\sigma} \int\limits_{-\infty}^{+\infty}
(t_{\bf k}(n)+E)J^{\gamma\sigma}_{\bf k}(E)dE.
\end{eqnarray}
Here
\begin{eqnarray} 
J^{\gamma\sigma}_{\bf k}(E)=\delta (E-E_{\gamma\sigma}({\bf k})) 
\theta (-E)
\end{eqnarray}
is the spectral intensity of Green function~(\ref{grf}), $\theta (-E)$ is the
step-wise function. 
From Eq.~(\ref{E0}) one can obtain for the 
ground state energy the expression
\begin{eqnarray}\label{E1} 
E_0=-{1\over 2} \sum_{\gamma\sigma} \left(\mu_{\gamma\sigma} n_{\gamma\sigma} 
-(1-\tau_1 n +\alpha_{\gamma\sigma}) n_{\gamma\sigma}
(1-n_{\gamma\sigma})w \right).
\end{eqnarray}
 
Expression~(\ref{E1})  can be rewritten in the form 
\begin{eqnarray}\label{E2} 
E_0&=&E^{(0)}_0+E^{(2)}_0+E^{(4)}_0,\\
E^{(0)}_0&=&{n\over 2}\left[-\mu+{n\over 4}\left(U+2U'-J_0-{zJ\over 8}
(16-(4-n)^2)\right)
-(1-\tau_1n-{n\over 2}\tau_2)(4-n){w\over 2}\right], \nonumber \\
E^{(2)}_0&=&\left( 2(1-\tau_1n-\tau_2(4-n))-{U\over w}-{J_0\over w}
+{2zJ\over w}(1+{n(4-n)\over 8})\right){w\over 8}m^2, \nonumber \\
E^{(4)}_0&=&{zJ\over 64}m^4. \nonumber 
\end{eqnarray}

The position of the mininum of ground state energy depends on 
values of model parameters. In Fig.~5 the energy difference $\Delta E_0$
between paramagnetic and ferromagnetic states is plotted as a function of 
the magnetization. At some
values of the parameters a ferromagnetic ordering with 
$m\neq 0$ is energetically gainful.
As it has been noted above the intra-atomic exchange is an important
factor leading to ferromagnetism in an orbitally degenerate band.
The increase of $J_0/w$ leads both to the increase of magnetic moment 
and to the decrease of the ferromagnetic ground 
state energy.

The dependence of the ground state energy of the model on the electron 
concentration is plotted in Fig.~6. One can see that with the increase of
$n$ Coulomb correlation becomes more and more important and the 
value of the ground state energy rapidly increases, as well as at the rise 
of the intra-atomic Coulomb repulsion parameters. 
As the value of intra-atomic exchange increases the  
ground state energy decreases (Fig.~7); at some critical value $J_0/w$ 
the transition of the system from a state of paramagnetic metal to a 
state of ferromagnetic metal occurs.
It appears that the ferromagnetic ordering can be more favourable
than the paramagnetic one in orbitally degenerate model without
singularities of the density of states even if the inter-atomic exchange 
and correlated hopping are absent (in the work~\cite{chao} similar result
is obtained only at the presence of density of states singularities). 
Let us also note that at increase of $U$ the critical value of $J_0$,
at which the transition to ferromagnetic state occurs, decreases. The
qualitatively similar picture has been obtained by the authors 
of work~\cite{klej}.

In Fig.~8 
the energy difference between the paramagnetic and ferromagnetic 
states (Fig.~8b) and 
the value of magnetization (Fig.~8a) as functions of band filling 
are plotted at different values of correlated hopping.
Depending on the value of $n$ (and the relation between the energy 
parameters) the state of the system can be para- or ferromagnetic, 
polarization can be full or partial.
The curves 1 correspond to the case when correlated hopping is absent.
With increase of $n$ the transition from para- to ferromagnetic state 
occurs, in the region $n>2$ the inverse transition takes place 
(symmetrical behaviour of the concentration dependence relative to
half-filling is observed).
The correlated hopping leads to the decrease of the ground state energy,
in particular, the increase of $\tau_1$ (curves 3) has stronger influence
at $n>2$  (as a result the region of ferromagnetic ordering moves towards
larger values of $n$),
the increase of $\tau_2$ (curves 2) -- at $n<2$ (the region of ferromagnetic 
ordering moves towards smaller values of $n$).
These effects are the manifestation of electron-hole asymmetry being
the property of the systems with correlated hoppings (see \cite{prb,cmp}).
At some values of correlated hopping in the system electron-hole symmetry
retrieves (curves 4 in Fig.~8).

The condition of ferromagnetic ordering stability 
${d^2 E_0 \over dm^2}<0$ can be obtained as 
\begin{eqnarray}
\label{umova}
{U+J_0\over 2w}+{zJ \over 8w}\left(8+n(4-n)\right)+n\tau_1
+{1\over 2}\tau_2(4-n)>1.
\end{eqnarray}
From expressions for ground state energy~(\ref{E2}) and  
magnetization~(\ref{m0})
one can see that
for the values of inter-atomic exchange interaction $J>0$
at the point of the transition from a paramagnetic metal to a ferromagnetic 
metal the magnetization changes continuously, and for $J=0$ it has a jump,
namely, in the former case the transition from
a paramagnetic state to a partially polarized ferromagnetic state occurs,
in the later one the transition from a paramagnetic state to a fully
polarized ferromagnetic state (saturated ferromagnetic state, $m=n$) 
is obtained.  The similar results have been obtained in a single-band 
model~\cite{hir_t0,iva,cm00}.
Thus, taking into account the inter-atomic exchange interaction
allows to obtain a partially polarized
ferromagnetic state in two-band Hubbard model with symmetrical density of 
states; the partially polarized ferromagnetic state has been obtained by 
the authors of work~\cite{oka}, using the special feature of the density of 
states.

For the case of $J=0$ from Eq.~(\ref{umova}) we obtain 
a generalization of the Stoner criterion which takes into account the 
orbital degeneracy and correlated hopping 
\begin{eqnarray}
(U+J_0)\rho(\epsilon_F) >1-n\tau_1-{1\over 2}\tau_2(4-n).
\end{eqnarray}
From the condition of the mininum of ground state energy ${d E_0 \over dm}=0$
one can also obtain the condition of partial spin polarization
\begin{eqnarray}
\label{umova_m}
{U+J_0\over 2w}+{zJ \over 8w}\left(8+n(4-n)-8m^2\right)+n\tau_1
+{1\over 2}\tau_2(4-n)>1.
\end{eqnarray}
The condition of full spin polarization ($m=n$) is
\begin{eqnarray}\label{fullsp}
{U+J_0\over 2w}+{zJ \over 8w}\left(8+n(4-n)-8n^2\right)+n\tau_1
+{1\over 2}\tau_2(4-n)>1.
\end{eqnarray}
Eqs.~(\ref{umova})-(\ref{fullsp}) coincide with the conditions which can be 
derived from Eq.~(\ref{m0}).  

From these conditions one can see that both mechanisms of correlated 
hopping favour ferromagnetism but their concentration dependences are
different;
if $\tau_2> \tau_1$ then the systems with the electron 
concentration $n<1$ are more favourable to ferromagnetism than the systems 
with $n>1$, and vice versa (see Figs.~2,~8,~9).

In Fig.~9  the critical values of inter-atomic exchange are plotted as a 
function of band filling. The lower curve corresponds to the critical value
for the partial polarization, the upper one -- for the full polarization;
the region below the lower curve corresponds to paramagnetic ordering of
spins, between the curves -- to the partial polarization, above the curves --
to the full polarization of spins. 
It can be seen that correlated hopping essentially changes the condition
of ferromagnetic ordering. Let us also note that the region of partial
polarization is narrowed with a deviation from half-filling (similarly to
the case of non-degenerate band~\cite{cm00}).
The maximum of critical value $zJ/w$ corresponds to almost empty band.
The cases $n=1,3$ are the most favourable for the existence of full spin 
polarization. Correlated hopping can lead to the displacement of the minimum 
points, and to 
the increase or decrease of $zJ/w$ critical values for the full 
polarization, i.e., to the non-equivalency of the cases $n<2$ and $n>2$.

In Fig.~10 the dependencies of critical values of $J_0/w$ on electron 
concentration at different values of $zJ/w$ are plotted.  
It is important to note that at $zJ=0$ the critical value of $J_0/w$ 
does not depend on the electron concentration.
It can be explained by the next arguments: in the absence of inter-atomic 
exchange the mechanism which stabilizes ferromagnetism is a translational 
motion of electrons which enforces the spins of sites, involved in the 
hopping process, to allign in parallel because of Hund's rule coupling.

Note also the essential difference of the situation where the system is 
described by non-zero values of correlated hopping: since correlated 
hopping renormalizes the bandwidth and makes it dependent on the 
concentration, the behaviour of the critical value 
of ${J_0 \over w}$ becomes asymmetrical relative to half-filling.

The critical value of $J_0/w$ as a function of $zJ/w$ is plotted on Fig.~11.
As one can see the increase of $zJ/w$ significantly decreases the critical
value of $J_0/w$ (in the same way the correlated hopping does).
It shows the importance of taking into account the inter-atomic exchange
and correlated hopping for the description of ferromagnetism in the systems
with orbital degeneracy. The inverse dependence of critical values of
$zJ/w$ and $J_0/w$ (what indicates the destabilization of ferromagnetic 
ordering at the increase of $J_0/w$) has been obtained in work~\cite{hir97} 
with use of the exact diagonalization method for the even number of sites
in one-dimensional chains, but that result depends sensitively on the number 
of lattice sites and the boundary conditions.

\section{Conclusions}
In this paper we have investigated the ground state of a doubly orbitally 
degenerate model. Taking into consideration the orbital degeneracy allows
to analyse the influence of intra-atomic exchange interaction (Hund's rule 
coupling), which is responsible for the formation of local magnetic moments, 
on the possibility of ferromagnetism realization.
Beside the diagonal matrix elements of electron-electron interactions 
the model includes the off-diagonal ones -- correlated hopping integrals,
which describe the influence of site occupancy on the hopping of electrons.  
The model under consideration also includes the inter-atomic exchange 
interaction  $J$. 

The study of the model ground state, carried out in this work, shows 
that the stability of ferromagnetism strongly depends on the model parameters.  
In particular, it has been found that the relationship between correlated 
hopping parameters determines the criterion of ferromagnetism. 
At $\tau_2> \tau_1$ in the system with concentration of electrons
$n<2$ the situation for ferromagnetic ordering is more favourable 
than for the system with  $n>2$; at $\tau_2< \tau_1$ the opposite
behaviour is obtianed. At some values of correlated hopping parameters
the retrieval of the electon-hole symmetry is possible.
Taking into account the correlated hopping leads to the appearance of 
the specific mechanism which stabilizes ferromagnetic ordering, and 
this is due to the spin-dependent shift of the subband centers.
In the absence of inter-atomic exchange interaction the ferromagnetic
ordering is stabilized by the translational motion of electrons between 
sites with ``atomic ferromagnetism'' formed by Hund's rule coupling.
  
It is important to note that the transition of the system from 
paramagnetic to ferromagnetic state can occur at the values
of interaction parameters, which are of the same order that bandwidth,
and with density of states without peculiarities.
The important role for the feromagnetism stabilization in weak interaction
regime ($U<2w$)
is played by the intra- and inter-atomic exchange interactions as well as 
correlated hopping that allows to describe the metallic
paramagnetic-ferromagentic transition with realistic relationship 
between above mentioned exchange interactions. 

For the values of inter-atomic exchange interaction $J>0$
at the point of the transition from a paramagnetic metal to a ferromagnetic 
metal the magnetization changes continuously, and for $J=0$ it has a jump,
namely, in the former case the transition from
a paramagnetic state to a partially polarized ferromagnetic state occurs,
in the later one the transition from a paramagnetic state to a fully
polarized ferromagnetic state (saturated ferromagnetic state) is obtained.  

The obtained dependencies of magnetization on concentration of electrons
qualitatively describe the experimental Slater-Pauling's curves for
ferromagnetic alloys. At some values of the model parameters 
the experimental dependence of magnetization for the systems   
Fe$_{1-x}$Co$_x$S$_2$ and Co$_{1-x}$Ni$_x$S$_2$ with changing
electron concentration in $e_g$ band is reproduced theoretically. 
The correlated hopping mechanism of ferromagnetism stabilization 
allows to explain the ferromagnetism in the systems Fe$_{1-x}$Co$_x$S$_2$
at small concentrations $x\simeq 0.05$.

\acknowledgments
V.H. is grateful to Prof. W.~Nolting (Humboldt-Universit\" at, Berlin)
for the hospitality during the workshop ``242. WE-Heraeus-Seminar on
Ground-State and Finite-Temperature Bandferromagnetism'' 
(4-6 October, 2000, Berlin), where the part of the results considered in
the present paper was discussed. The authors thank to Prof. D.~Vollhardt 
and Prof. W.~Weber for valuable discussions.

\newpage
\begin{figure}[hpb] 
\begin{minipage}[b]{75mm}
\epsfxsize=75mm
\epsfysize=65mm
\epsfclipon
\epsffile{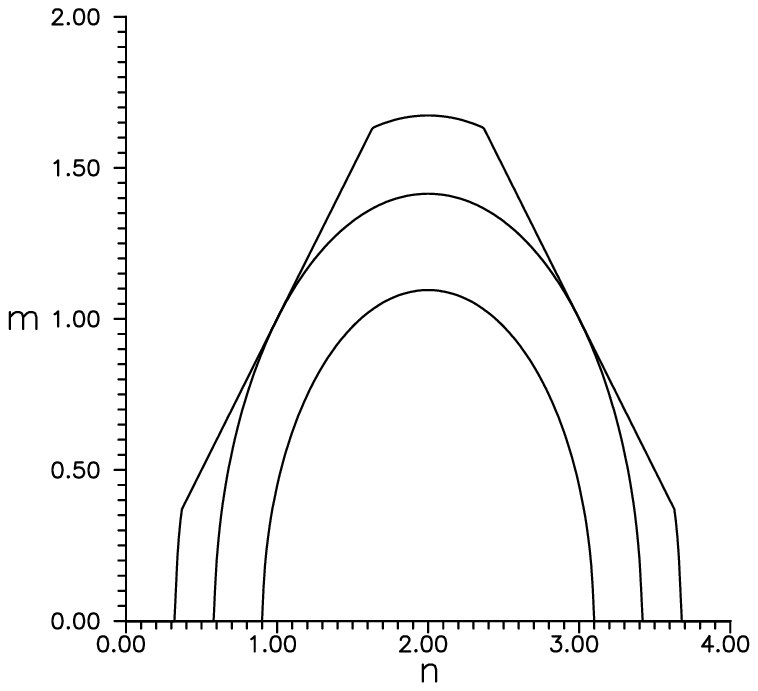}
\renewcommand{\baselinestretch}{1}
\caption{The magnetization $m$ as a function of 
$n$ at ${U/w}=1.5$ and $\tau_1=\tau_2=0$,
${zJ/w}=0.1$.
Upper curve corresponds to ${J_0/w}=0.27$, middle curve
corresponds to ${J_0/w}=0.25$, lower one 
correponds to ${J_0/w}=0.23$.}
\end{minipage}
\hfill
\begin{minipage}[b]{75mm}
\epsfxsize=75mm
\epsfysize=65mm
\epsfclipon
\epsffile{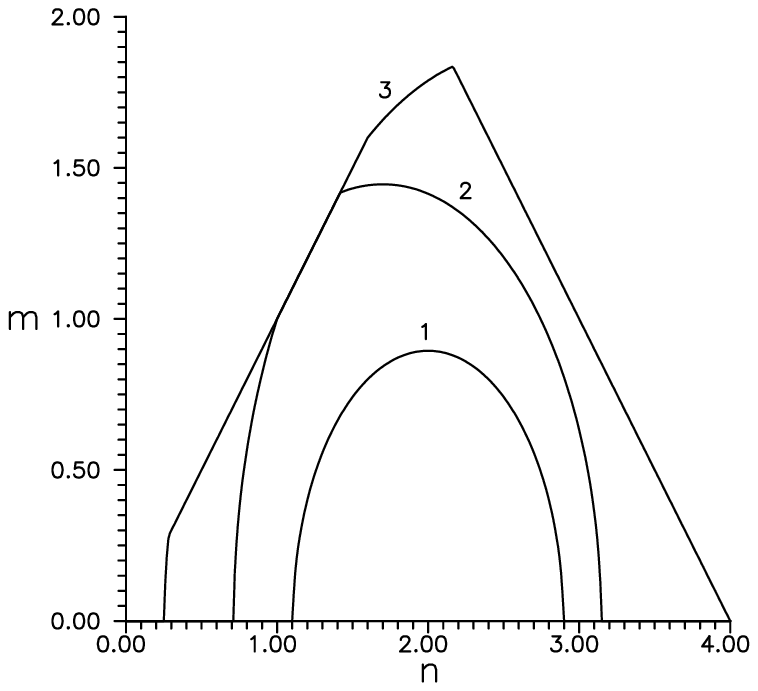}
\renewcommand{\baselinestretch}{1}
\caption{The magnetization $m$ as a function of 
$n$ at $U/w=1.5$, $zJ/w=0.1$ 
and $J_0/w=0.22$.
Curve 1 corresponds to $\tau_1=\tau_2=0$,
curve 2 -- to $\tau_1=0$, $\tau_2=0.015$
curve 3 -- to $\tau_1=0.015$, $\tau_2=0$.}
\end{minipage}
\vfill	
\begin{minipage}[t]{75mm}
\epsfxsize=75mm
\epsfysize=65mm
\epsfclipon
\epsffile{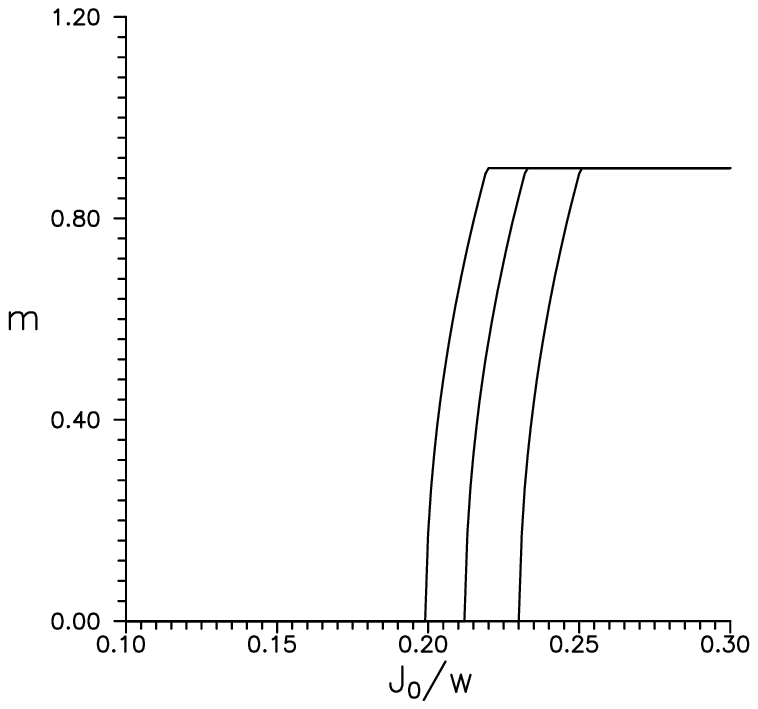}
\renewcommand{\baselinestretch}{1}
\caption{The magnetization $m$ as a function of 
$J_0/w$ at $n=0.9$, $U/w=1.5$ 
and $zJ/w=0.1$.
Left curve corresponds to $\tau_1=0$, $\tau_2=0.01$, middle curve
corresponds to $\tau_1=0.01$, $\tau_2=0$, right one 
correponds to $\tau_1=\tau_2=0$.}
\end{minipage}
\hfill
\begin{minipage}[t]{75mm}
\epsfxsize=75mm
\epsfysize=65mm
\epsfclipon
\epsffile{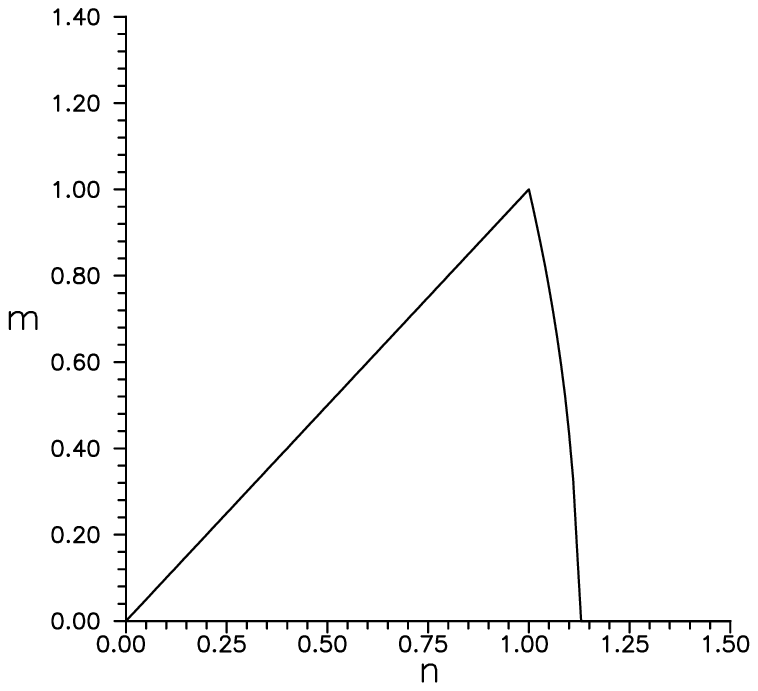}
\renewcommand{\baselinestretch}{1}
\caption{The magnetization $m$ as a function of 
$n$ at $U/w=1.2$, $zJ/w=0.06$ 
and $J_0/w=0.2$, $\tau_1=0$, $\tau_2=0.15$.}
\end{minipage}
\end{figure}

\begin{figure}[hpbt] 
\begin{minipage}[t]{75mm}
\epsfxsize=75mm
\epsfysize=65mm
\epsfclipon
\epsffile{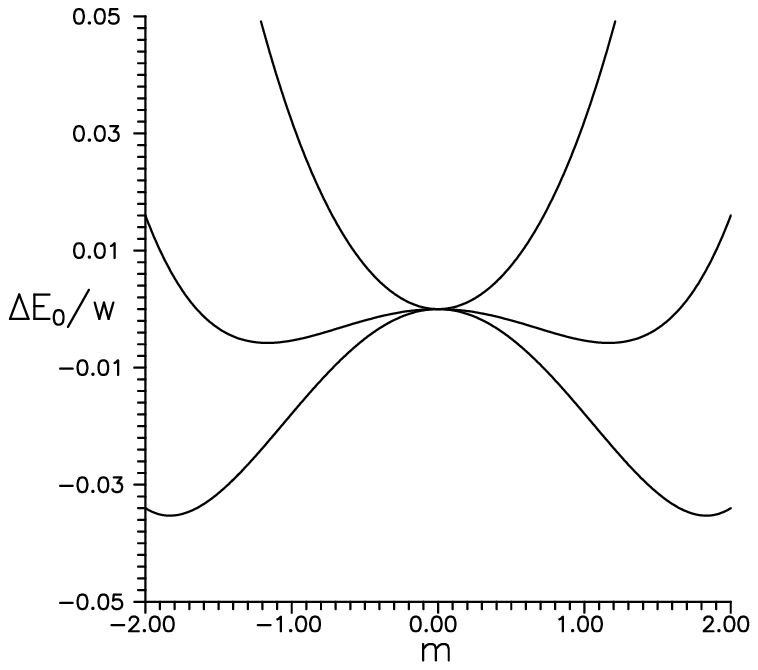}
\renewcommand{\baselinestretch}{1}
\caption{The energy difference between ferro- and paramagnetic ground
states  as a function of 
magnetization $m$  at $n=1.2$, ${U/w}=1.2$, 
${zJ/w}=0.2$ and $\tau_1=\tau_2=0$:
upper curve corresponds to ${J_0/w}=0$,
middle curve corresponds to ${J_0/w}=0.3$
and lower curve corresponds to ${J_0/w}=0.4$.}
\end{minipage}
\hfill
\begin{minipage}[t]{75mm}
\epsfxsize=75mm
\epsfysize=65mm
\epsfclipon
\epsffile{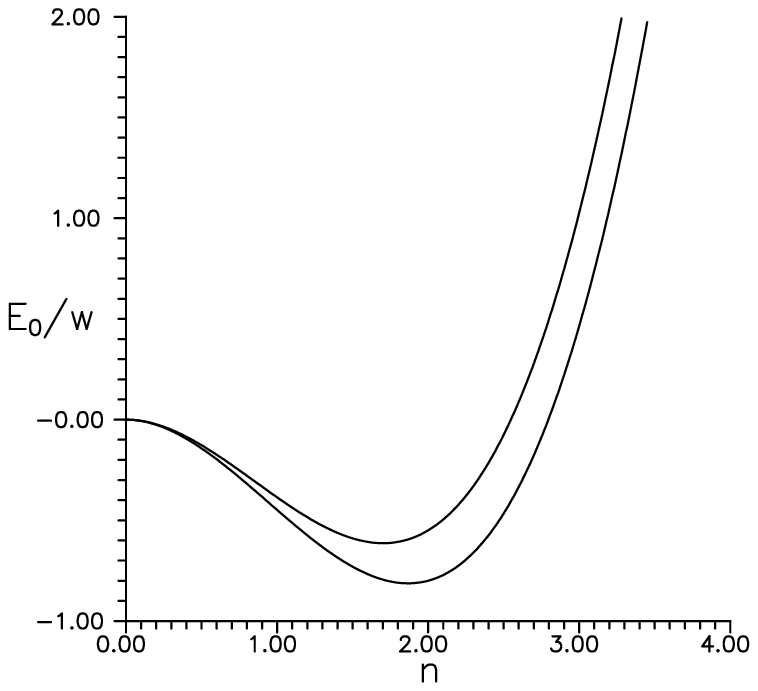}
\renewcommand{\baselinestretch}{1}
\caption{The energy of ferromagnetic ground state as a function
of $n$ at $U/w=1.2$, $zJ/w=0.1$, 
$\tau_1=\tau_2=0$: $J_0/w=0.1$ for upper curve,
$J_0/w=0.2$ for lower one.}
\end{minipage}
\vfill
\begin{minipage}[t]{75mm}
\epsfxsize=75mm
\epsfysize=65mm
\epsfclipon
\epsffile{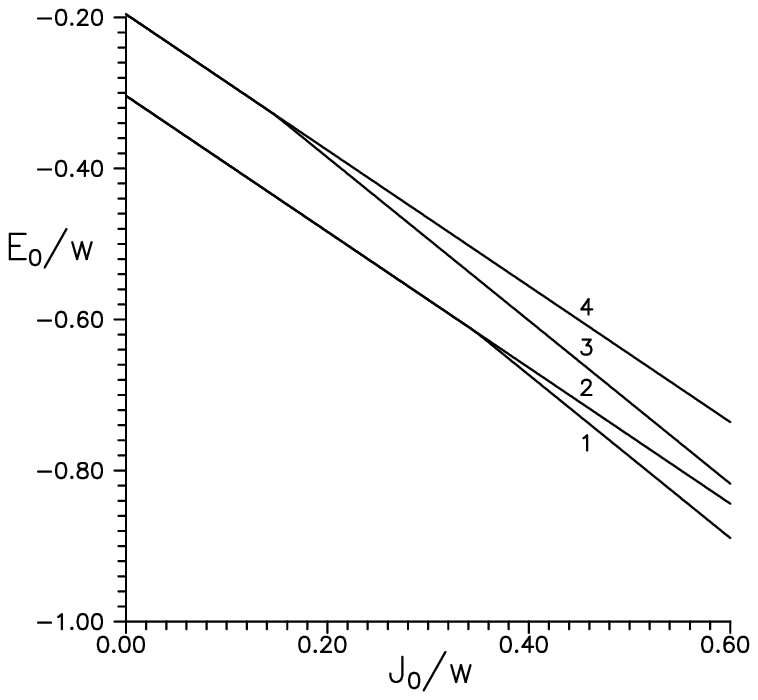}
\renewcommand{\baselinestretch}{1}
\caption{
The ferromagnetic (curves 1,3)
and paramagnetic (curves 2,4)
ground state energies as a function of 
$J_0/w$: $n=1.2$, $zJ/w=0.05$,
$\tau_1=\tau_2=0.1$;
curves 1,2 correspond to  $U/w=1$,
curves 3,4 correspond to  $U/w=1.2$.}
\end{minipage}
\end{figure}

\newpage
\begin{figure}[htb]
\begin{minipage}{100mm}
\epsfxsize=75mm
\epsfysize=65mm
\epsfclipon
\epsffile{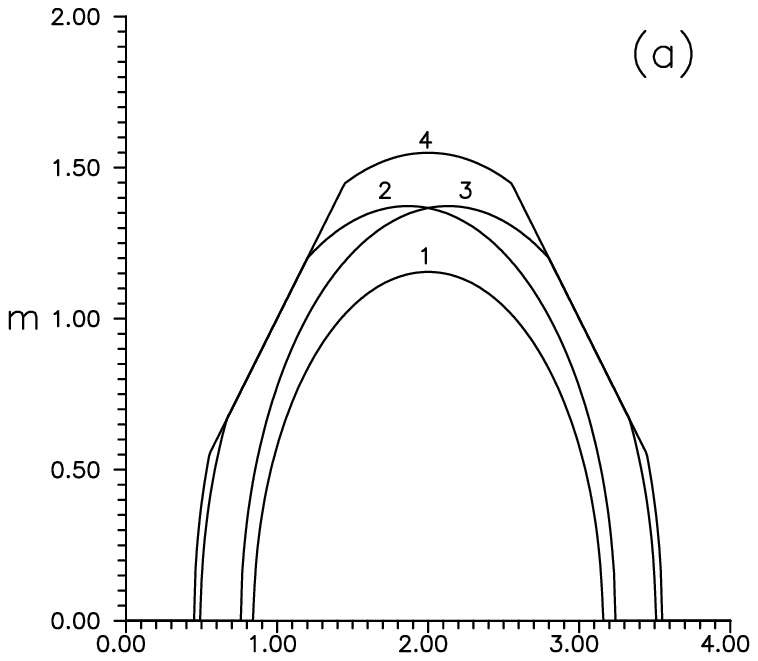}
\end{minipage}

\begin{minipage}{100mm}
\epsfxsize=75mm
\epsfysize=65mm
\epsfclipon
\epsffile{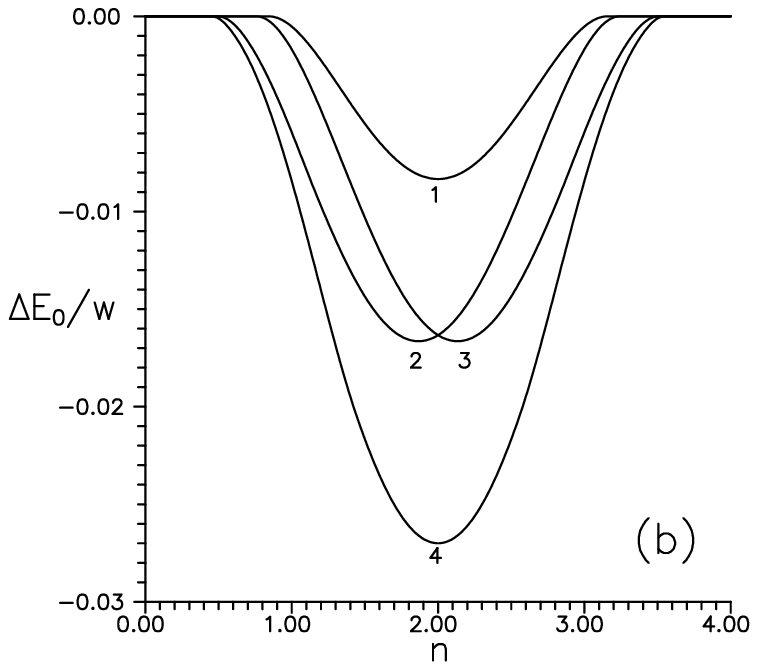}
\end{minipage}
\renewcommand{\baselinestretch}{1}
\caption{The magnetization $m$ (panel a)
and the energy difference between ferro- and paramagnetic ground
states (panel b)
as a function of $n$ at $U/w=0.9$, $zJ/w=0.3$, 
$J_0/w=0.3$:
curve 1 correspond to  $\tau_1=\tau_2=0$,
curve 2 - to $\tau_1=0$, $\tau_2=0.02$,
curve 3 - to $\tau_1=0.01$, $\tau_2=0$
and curves 4 to $\tau_1=0.01$, $\tau_2=0.02$.}
\end{figure}

\newpage
\begin{figure}[hpb] 
\begin{minipage}[b]{75mm}
\epsfxsize=75mm
\epsfysize=65mm
\epsfclipon
\epsffile{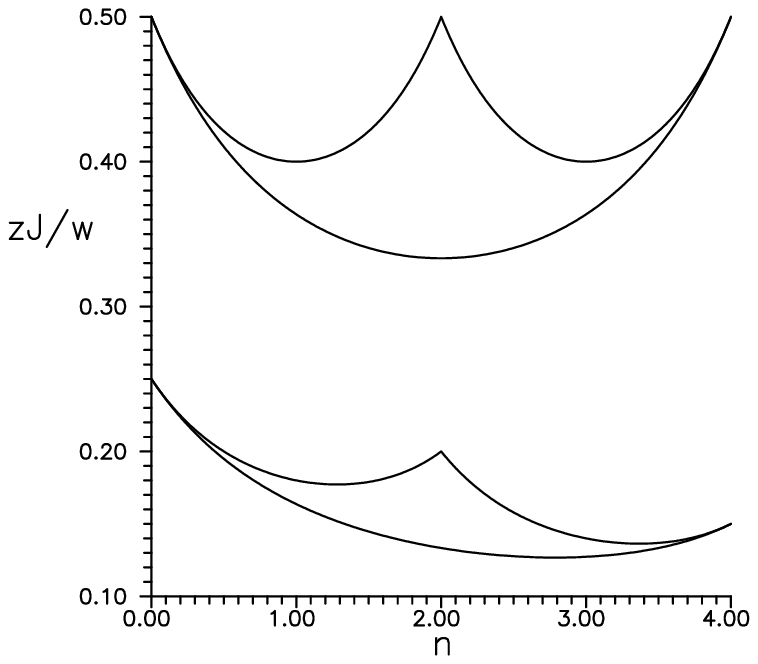}
\renewcommand{\baselinestretch}{1}
\caption{The critical values of $zJ/w$
as a function of $n$ at $J_0/w=0.1$.
Upper curves correspond to $U/w=0.9$, 
$\tau_1=\tau_2=0$,
lower curves  - to $U/w=1.2$,
$\tau_1=0.05$.}
\end{minipage}
\hfill
\begin{minipage}[b]{75mm}
\epsfxsize=75mm
\epsfysize=65mm
\epsfclipon
\epsffile{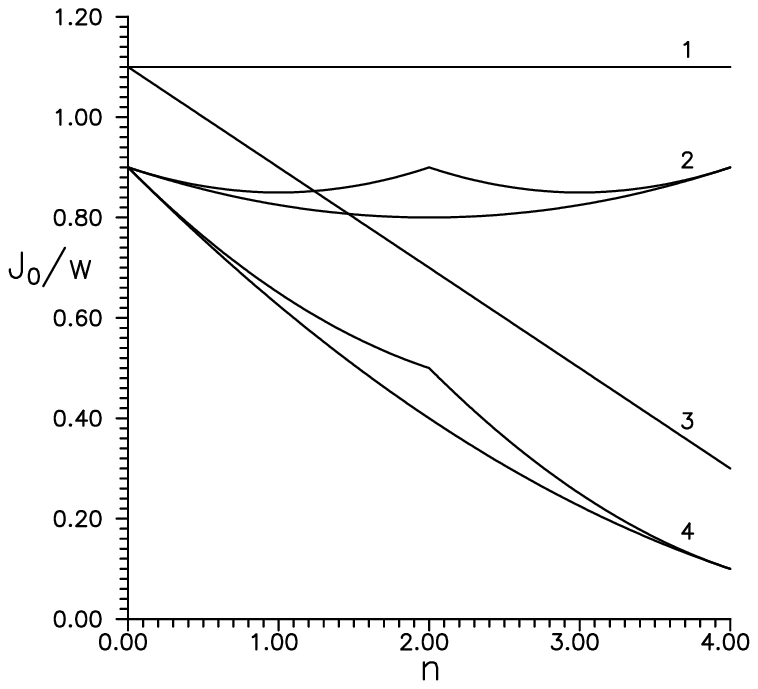}
\renewcommand{\baselinestretch}{1}
\caption{The critical values of $J_0/w$
as a function of 
$n$ at $U/w=0.9$, $\tau_2=0$.
Curve 1 corresponds to $zJ/w=0$, $\tau_1=0$,
curves 2 - to $zJ/w=0.1$, $\tau_1=0$, 
curve 3 - to $zJ/w=0$, $\tau_1=0.01$, 
and curves 4 correspond to $zJ/w=0.1$, $\tau_1=0.01$.}
\end{minipage}
\vfill
\begin{minipage}[b]{75mm}
\epsfxsize=75mm
\epsfysize=65mm
\epsfclipon
\epsffile{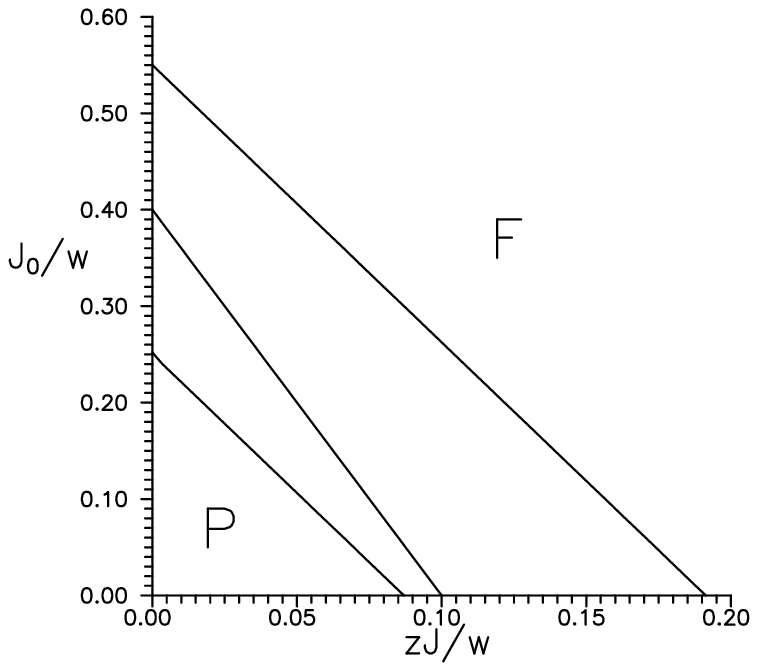}
\renewcommand{\baselinestretch}{1}
\caption{The critical value of ${J_0/ w}$ vs ${zJ/ w}$
at ${U / w}=1$ and $\tau_1=\tau_2=0.1$
Upper  curve  corresponds to $n=0.5$,
middle curve  - to $n=2$,
lower curve - to $n=3.5$.}
\end{minipage}
\end{figure}

\end{document}